\documentclass{vgtc}         

\graphicspath{{figures/}{pictures/}{images/}{./}} 

\usepackage{times}                     

\usepackage{tabu}                      
\usepackage{booktabs}                  
\usepackage{lipsum}                    
\usepackage{mwe}                       
\usepackage{url}
\usepackage{amssymb,amsmath}
\usepackage{tabularx} 
\usepackage{booktabs}

\usepackage{amsmath,amssymb,graphicx,braket,caption,subcaption,braket,qcircuit,bm}
\DeclareMathOperator{\Tr}{Tr} 
\DeclareMathOperator{\Real}{Re} 
\DeclareMathOperator{\Imag}{Im} 

\usepackage{booktabs}
\usepackage{makecell} 
\usepackage{caption} 

\onlineid{0}

\vgtccategory{Research}

\vgtcinsertpkg

\title{Quantum Intuition XR: Tangible Quantum Mechanics using Interactive XR Experience}

\author{Jamie Ngoc Dinh\thanks{e-mail: ngocdinh@umd.edu} %
\and Marven Wong\thanks{e-mail: marven@terpmail.umd.edu} %
\and Matthew Brooks\thanks{e-mail: mbrook11@umd.edu}
\and Charles Tahan\thanks{e-mail: ctahan@umd.edu}
\and Myungin Lee\thanks{e-mail: myungin@umd.edu}}
\affiliation{\scriptsize University of Maryland College Park}

\teaser{
  \centering
  \includegraphics[width=\linewidth]{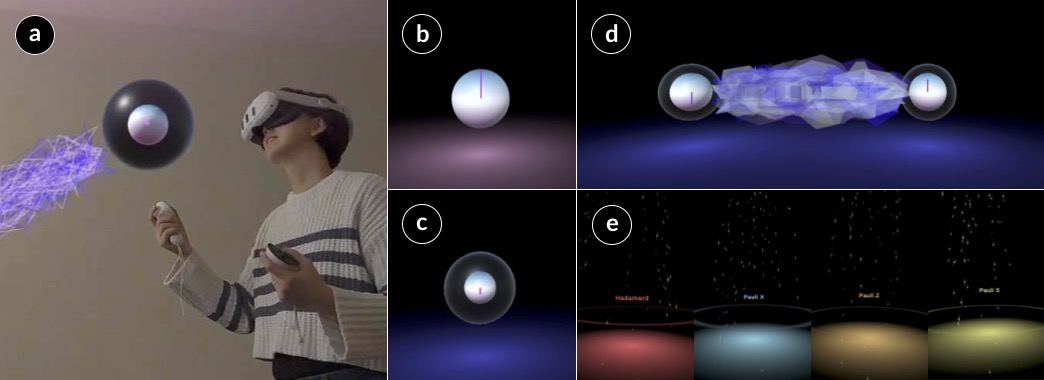}
  \caption{(a) A user holding a qubit in an Augmented Reality environment. (b) A qubit in its unentangled state represented by the well-known Bloch sphere. (c) A qubit in its entangled state, where the size of the Bloch sphere accurately reflects the magnitude of entanglement with other qubits (and shrinks to zero at full entanglement).  (d) Two entangled qubits where the entanglement between two qubits is depicted by an arc--varying by entanglement entropy. (e) Visualization of four quantum gates-Hadamard, Pauli-X, Pauli-Y, Pauli-Z.}
  \label{fig:rep}
}

\abstract{
    Understanding quantum mechanics is inherently challenging due to its counterintuitive principles. Quantum Intuition XR is an interactive, extended reality (XR) experience designed to make quantum concepts tangible. Our system visualizes core principles of quantum computing, including qubits, superposition, entanglement, and measurement, through immersive interaction. Using a Mixed Reality headset, participants engage with floating qubits, manipulate their states via controllers, and observe entanglement dynamics through real-time audiovisual feedback. A key feature of our implementation is the mathematically accurate and dynamic representation of qubits, both individually and while interacting with each other. The visualization of the qubit states evolve---rotate, shrink, grow, entangle---depending on their actual quantum states, which depend on variables such as proximity to other qubits and user interaction. Preliminary expert interviews and demonstrations with quantum specialists indicate that the system accurately represents quantum phenomena, suggesting strong potential to educate and enhance quantum intuition for non-expert audiences. This approach bridges abstract quantum mechanics with embodied learning, offering an intuitive and accessible way for users to explore quantum phenomena. Future work will focus on expanding multi-user interactions and refining the fidelity of quantum state visualizations.} 


\begin{document}
\maketitle

\section{Introduction}
The rules that describe how the universe works at small and low energy scales--\textit{quantum mechanics}--are notoriously counter-intuitive as they lack correspondence with everyday experience. Unlike classical physics, quantum phenomena such as superposition and entanglement are abstract and difficult to visualize, which contributes to persistent learning challenges. Freericks et al. (2019)~\cite{freericks_teaching_2019} emphasize the value of concept-focused instruction supported by multimedia tools to make quantum ideas more accessible. Bouchée et al. (2021)~\cite{bouchee_investigating_2023} similarly report that students and teachers benefit from visual aids and contextual applications to support understanding.

Traditional computer-generated visualizations, such as videos, animations and simulations, are widely used to represent complex and abstract processes in quantum mechanics~\cite{kohnle_new_2012}~\cite{ahmed_quantum_2020}. While there is evidence that such dynamic visualizations can enhance understanding of such abstract concepts~\cite{hoffler_instructional_2007}, other studies have found that they are not always more effective than static images~\cite{tversky_animation_2002}. Additionally, they remain limited by their presentation on 2D screens and passive role they assign to the users. Koning and Tabber (2011)~\cite{de_koning_facilitating_2011} suggests that the effectiveness of dynamic visualizations can be improved by grounding them in bodily experience - such as gesture, interaction, and spatial movement - highlighting the potential of immersive technologies like Extended Reality to support embodied learning. 

Extended Reality (XR), which merges digital content with the physical environment, has expanded the potential for immersive and interactive experiences. By enveloping users in spatially coherent environments that allow moving and manipulating virtual objects, XR enables a sense of presence and immersion that goes beyond traditional screen-based media. Considering quantum mechanics is not normally visible or audible in real life, providing an interactive immersive experience using XR can assist broader audiences in understanding the quantum mechanics through hands-on experience.

This project, titled "Quantum Intuition XR (QIXR)", aims to make quantum mechanics more tangible by leveraging XR as the medium. It integrates core quantum principles into the audio, visual, and interactive elements of an immersive extended reality environment, enabling users to engage with abstract concepts through embodied, multisensory experiences. The project is particularly inspired by the call to action in A Quantum Wish~\cite{staff_quantum_2022}, which emphasizes the need to advance quantum education in order to foster a skilled, diverse, and inclusive workforce for the quantum future.



The XR experience presents users with two interactive qubits floating in space, surrounded by four quantum gates-Hadamard ($H$), Pauli-X ($X$), Pauli-Z ($Z$), and Phase-S ($S$). Each qubit is visually rendered to display its current quantum state using Bloch sphere, allowing real-time observation of quantum behavior. Users can grab and manipulate the qubits within the virtual environment; placing a qubit into one of the gates applies a corresponding transformation to its Bloch sphere. When the two qubits are brought into close proximity, an exchange interaction—such as entanglement—will be triggered, dynamically altering their states. This interface is designed to foster an embodied and intuitive understanding of quantum mechanics by enabling hands-on exploration of complex phenomena through spatial interaction. 


In this early stage, the system supports a single-player virtual reality experience. Future developments will focus on expanding this to an Augmented Reality setting using passthrough technology and enabling multi-user interactions. Initial steps towards AR implementation have been successfully achieved, incorporating hand controls through OpenXR, the XR Interaction Toolkit, and the AR Foundation package. In summary, our contributions are as follows:
\begin{itemize}
    \item Design and implementation of QIXR, an XR system that visualizes quantum concepts through real-time, interactive audio-visual feedback. 
    \item An embodied interface for intuitive exploration, enabling users to manipulate qubits, apply gates, and trigger entanglement via spatial interaction.
    \item Technical validation and preliminary expert interviews, demonstrating the system’s potential to support quantum education for non-experts and laying the groundwork for future multi-user AR extensions.
\end{itemize}

\section{Related Work} 




\subsection{Visualization of Quantum Mechanics}
A variety of computer-based visualizations—including videos, animations, and interactive simulations—have long been used to convey the complex and abstract principles of quantum mechanics. The QuVis project~\cite{kohnle_new_2012} offers a collection of interactive animations that guide learners through key quantum concepts using step-by-step explanations and targeted visual cues. Quantum Composer~\cite{ahmed_quantum_2020}, on the other hand, provides a modular, node-based interface that allows users to build and manipulate quantum systems, promoting exploratory and customizable learning. Migdal et al. (2022)~\cite{migdal_visualizing_2022} introduced Virtual Lab, a browser-based interactive simulation platform that visualizes quantum phenomena—including entanglement, measurement, and quantum computing—through an intuitive drag-and-drop interface, enabling users to explore quantum mechanics in a no-code, visually rich environment. These tools represent important progress in making quantum concepts more accessible, yet they remain bound to 2D screens and often lack the embodied engagement necessary for deeper intuitive understanding.

\subsection{Science in XR}
Extended Reality (XR), which encompasses Virtual Reality (VR), Augmented Reality (AR), and Mixed Reality (MR), is playing an increasingly significant role in scientific simulations and science education~\cite{draschkow_using_2023}. A key advantage of XR lies in its ability to model real-world phenomena that are too small, large, hazardous, or complex to observe directly, while also allowing users to interact with virtual objects that supports deeper understanding and engagement. For instance, Interactive Molecular Dynamics in Virtual Reality (iMD-VR)~\cite{crossley-lewis_interactive_2023} enables real-time manipulation of molecular structures, providing an intuitive interface for exploring molecular interactions. Similarly, Norrby et al. (2015)~\cite{norrby_molecular_2015} introduced Molecular Rift, a gesture-controlled VR system for visualizing and interacting with 3D molecular models. These immersive tools offer unique benefits for drug discovery, materials design, and science education that go beyond the capabilities of traditional 2D visualizations. Reen et al. (2022)~\cite{reen_developing_2022} developed a student co-designed immersive VR simulation for teaching complex molecular biology concepts, demonstrating that spatial interaction significantly enhanced students’ understanding, engagement, and retention of abstract biological processes. In the domain of physics, Kokiadis et al. (2024) proposed a decoupled physics architecture for XR applications, using edge-cloud infrastructure to offload computation and improve performance~\cite{kokiadis_decoupled_2024}.

\subsection{Quantum Mechanics in XR}
The integration of quantum mechanics with extended reality (XR) has gained traction as a means to enhance understanding of complex quantum phenomena. Li~\cite{li_simulating_2023} introduced an AR-based Quantum Turing Machine simulation, enabling users to visualize quantum computation interactively. Similarly, Tarng and Pei~\cite{tarng_application_2023} demonstrated that VR-based quantum learning environments improve comprehension of abstract concepts like superposition and entanglement. Zable et al.~\cite{zable_investigating_2020} developed interactive VR modules for experimenting with quantum circuits, reinforcing quantum logic through direct manipulation. Another relevant study by Sisini et al.~\cite{sisini_quantum_2022} introduced a 3D virtual environment using CoSpaces for teaching quantum computing fundamentals, providing students with an interactive quantum laboratory. Weymuth and Reiher~\cite{weymuth_immersive_2020} extended these ideas to the field of quantum chemistry, employing VR and haptic feedback systems to facilitate an intuitive understanding of quantum interactions in chemical reactions. These works demonstrate that XR can bridge the gap between abstract quantum theory and intuitive learning experiences.

\section{Quantum Intuition: Principle and Implementation} 

QIXR aims to make quantum mechanics tangible by simulating scientifically accurate quantum behaviors within virtual environment. In the proposed system, we specifically simulated the following concepts: qubits and the Bloch sphere, exchange interaction, quantum gates, entanglement, and measurements. In the following, we introduce these core concepts based on ~\cite{griffiths2018introduction, nielsen_quantum_2002} and corresponding implementations in Unity3D game engine~\cite{noauthor_unity_nodate}.

\begin{figure*} [ht]
    \centering
    \includegraphics[width=\linewidth]{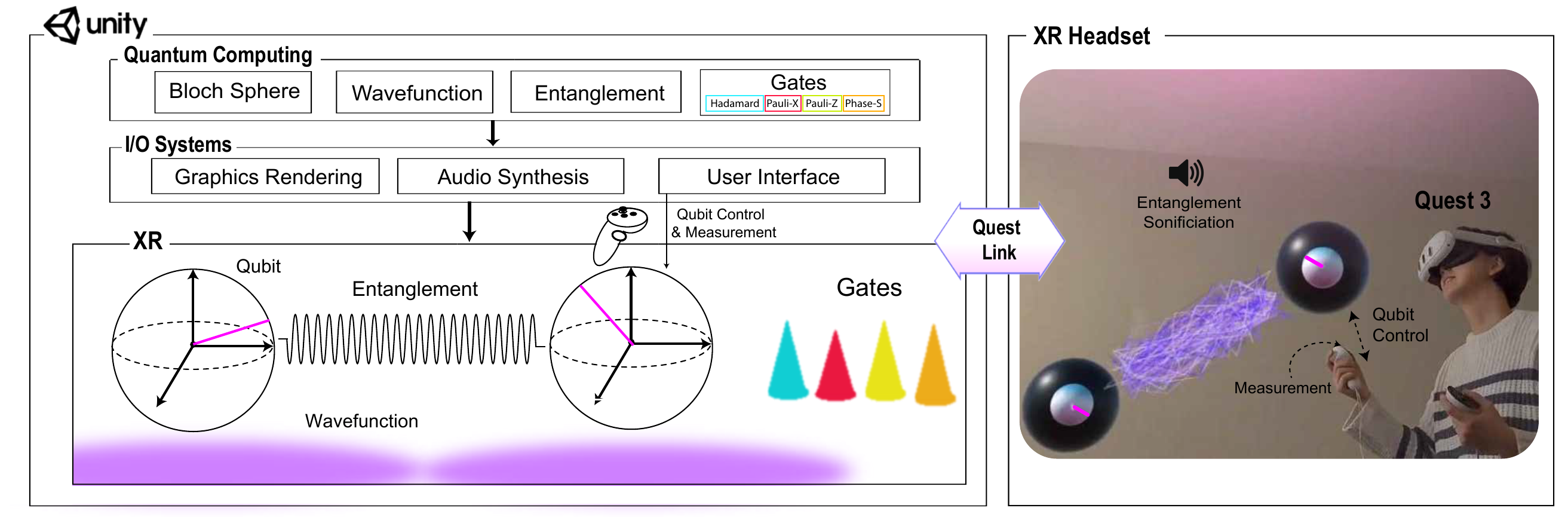}
    \caption{The schematic of the QIXR. The proposed system comprises Quantum computing, Input/Output (I/O) Systems, and XR. Quantum Computing includes Bloch sphere, Wavefunction, Entanglement, and Gates. I/O connects the XR domain with Quantum Computing with graphic rendering, audio synthesis, and user interface. The XR scene created on the PC is linked (wireless or wired) to the XR headsets for user experience.}
    \label{fig:schematic}
\end{figure*}

For implementation, the system utilizes Meta Quest 3 device in which Unity3D game engine \cite{noauthor_unity_nodate} and programming language C\# is used to develop the application. Using Oculus Link, the headset wirelessly streams stereo audio and 3D visuals from Unity3D, allowing users to naturally interact with the audio-visual stimulation of quantum mechanics. Figure~\ref{fig:schematic} presents a schematic of our system, illustrating the workflow from quantum computing in Unity3D to the user interface in the Meta Quest headset. 

\subsection{Qubits and the Bloch Sphere}
In nature, the behavior of some objects, usually very small or cold (low energy) objects, is governed by quantum mechanics. Simply put, a quantum object has observables, physical measurable quantities such as energy or momentum, that are quantized, i.e. they are locked to specific values. This is in contrast to the conventional behavior of the world around us, referred to as classical mechanics. A classical object, say a car for example, have continuous observables, i.e. the car can travel at any speed as opposed just a few fixed speeds. This counter-intuitive behavior of quantum objects may be technologically harnessed with extraordinary consequences. One such technology, is a quantum computer.

The fundamental building block of a computer is a bit, an electronic component known as a transistor that carries the binary information 0 or 1. Similarly, a quantum computers fundamental building block is a quantum-bit or a \textit{qubit}. A qubit is a quantum two-level object that also encodes binary information. However, a consequence of a quantum objects discreet behavior is that it is also probabilistic. This means that, until an observable, which in the case of a qubit is the binary information, is measured, it exists in a state that has a probability of being measured as both possible outcomes 0 and 1. This is known as a \textit{superposition}.

The superposition state of a qubit is described with a vector using the \textit{bra-ket} notation. In this notation, the binary information (classical) states are given by the two orthogonal column vectors or kets
\begin{equation}
    \ket{0}=\begin{pmatrix}
           1\\
           0
         \end{pmatrix}\quad\quad
         \ket{1}=\begin{pmatrix}
           0\\
           1
         \end{pmatrix}.
\end{equation}

\noindent A qubits superposition state $\ket{\psi}$ may therefore be written as

\begin{equation}
    \ket{\psi}=\alpha \ket{0} + \beta \ket{1}=\begin{pmatrix}
           \alpha\\
           \beta
         \end{pmatrix}
\end{equation}

\noindent where $\alpha$ and $\beta$ are complex numbers that obey the \textit{normalization condition}

\begin{equation}
    |\alpha|^2+|\beta|^2 = 1.
\end{equation}

\noindent The normalization condition guarantees that when the qubit is measured/observed it will only ever be measured as a $\ket{0}$ or a $\ket{1}$.

\begin{figure}[t] 
    \centering
    \includegraphics[width=1\linewidth]{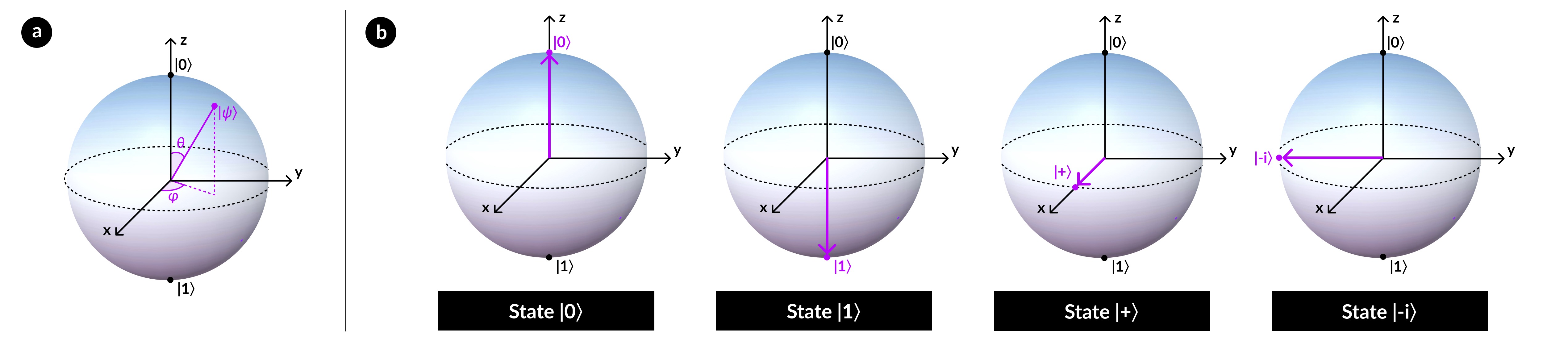}
    \caption{(a) The Bloch sphere that provides visualization of the state of a qubit. (b) Examples of visualizations of qubits with different spins.}
    \label{fig:qubitblochsphere}
\end{figure}

Another general way of writing the superposition of a qubit is as 

\begin{equation}
    \ket{\psi}=\cos\left(\frac{\theta}{2}\right) \ket{0} + e^{i \phi}\sin\left(\frac{\theta}{2}\right) \ket{1}
    \label{eq:psi}
\end{equation}

\noindent where the $\theta$ and $\phi$ are real numbers in the ranges $0\leq\theta\leq \pi$ and $0\leq\phi\leq 2 \pi$. This description is useful because it naturally denotes the superposition of a qubit as a point on the \textit{Bloch sphere}. The Bloch sphere is a unit sphere, meaning it has radius $r=\sqrt{x^2+y^2+z^2}=1$ arbitrary units, and the two angles $\theta=\arccos(z/\sqrt{x^2+y^2+z^2})$ and $\phi=(y/|y|)\arccos(x/\sqrt{x^2+y^2})$ describe every point on the surface of the sphere. This is shown in Figure~\ref{fig:qubitblochsphere}.

\subsubsection{Density Matrix}
Another useful description of the state of a qubit is as a \textit{density matrix}. The density matrix of a single qubit is given as

\begin{equation}
    \rho=\ket{\psi}\bra{\psi}=\begin{pmatrix}
    \rho_{00} & \rho_{01}\\
    \rho_{10} & \rho_{11}
    \end{pmatrix}
    \label{eq:Single_Density_Matrix}
\end{equation}

\noindent where the $\bra{\psi}$ is the conjugate transpose of the state vector $\ket{\psi}$, given by a row vector or bra. For example, for the single qubit state described in~\ref{eq:psi}, the corresponding density matrix is

\begin{equation}
\begin{split}
        \rho=\ket{\psi}\bra{\psi}&=\begin{pmatrix}
           \cos\left(\frac{\theta}{2}\right)\\
           e^{i \phi}\sin\left(\frac{\theta}{2}\right)
         \end{pmatrix}.\left(\cos\left(\frac{\theta}{2}\right),e^{-i \phi}\sin\left(\frac{\theta}{2}\right)\right)\\
         &=\begin{pmatrix}
    \cos\left(\frac{\theta}{2}\right)^2 & \frac{1}{2}e^{-i \phi} \sin\left(\theta\right)\\
    \frac{1}{2}e^{i \phi} \sin\left(\theta\right) & \sin\left(\frac{\theta}{2}\right)^2 
    \end{pmatrix}.
\end{split}
\label{eq:Psi_Density_Matrix}
\end{equation}

\noindent For a density matrix, the normalization condition is that the sum of the diagonal elements of the matrix, known as the \textit{trace}, is 1. This is evident for~\ref{eq:Psi_Density_Matrix} as $\Tr [ \rho ]=\cos\left(\frac{\theta}{2}\right)^2+\sin\left(\frac{\theta}{2}\right)^2 = 1$. The state described by a density matrix may be translated onto a Bloch sphere by

\begin{subequations}
    \begin{equation}
        u=\rho_{01}+\rho_{10}=2\Real(\rho_{01})
    \end{equation}
    
    \begin{equation}
        v=i(\rho_{01}-\rho_{10})=2\Imag(\rho_{01})
    \end{equation}

    \begin{equation}
        w=\rho_{00}-\rho_{11}
    \end{equation}
    \label{eq:Cartesian_Density_Matrix}
\end{subequations}

\noindent where $\{u,v,w\}$ are the Cartesian coordinates of the state. Although the density matrix representation may seem unnecessary when discussing a single qubit, as all dynamics of the qubit can be sufficiently described as a state vector as in~\ref{eq:psi}, when describing the dynamics of two or more interacting qubits, the density matrix representation is essential.

\subsubsection{Unity Implementation}

To simulate qubits within our Unity scene, Math.NET is utilized for its ability to represent complex 32 bit numbers $\langle r,i \rangle$, where r is the real component and i is the imaginary component. Each qubit is represented as a $2\times2$ density matrix $\rho$ of complex 32 bit numbers
\begin{equation}
 \rho = 
    \begin{pmatrix}
           \langle 1,0 \rangle & \langle 0,0 \rangle \\ 
           \langle 0,0 \rangle & \langle 0,0 \rangle
    \end{pmatrix}
    \label{eq:densitymatrix}
\end{equation}

\noindent To visually depict the qubit, a qubit's 2x2 matrix is represented as the Bloch sphere with polar coordinates using complex 32 bit numbers. By taking the Bloch vector (u,v,w), we can utilize these points as (x,y,z) in Unity's coordinate plane of Cartesian coordinates. 


\begin{equation}
 \rho = 
    \begin{pmatrix}
           \langle r_{00},i_{00} \rangle & \langle r_{01},i_{01} \rangle \\ 
           \langle r_{10},i_{10} \rangle & \langle r_{11},i_{11} \rangle
    \end{pmatrix}
\end{equation}

\begin{equation}
 u = 2 \langle r_{01} \rangle
\end{equation}
\begin{equation}
 v = 2 \langle i_{10} \rangle
\end{equation}
\begin{equation}
 w = \langle r_{00},i_{00} \rangle - \langle r_{11},i_{11} \rangle.
\end{equation}

The qubits are stored together in a $2^n$ by $2^n$ density matrix where $n$ is the number of qubits simulated within the scene in Unity. As more qubits are added within the scene, the size of the density matrix grows exponentially to accommodate new qubits.

\subsection{Quantum Gates}
In classical computers, logic gates are used to perform the calculations that drive our day-to-day uses. Typically, logic gates will take two input bits and output a separate bit depending on the values of those input bits and the type of gate. Generally, after a classical logic gate has been performed, it cannot be undone, as for most logic gates the input bit values cannot be determined from the output bit value. This is not the case for a quantum computer.

Quantum gates, the quantum equivalent of logic gates are written as matrices of size $2^n \times 2^n$ where $n$ is the number of qubits that the gate affects. Some examples of gates that act on only one qubit (single-qubit gates) are 

\begin{equation}
\begin{split}
    Z=\begin{pmatrix}
           1 & 0\\ 
           0 & -1
         \end{pmatrix},\quad\quad\quad
    X=\begin{pmatrix}
           0 & 1\\ 
           1 & 0
         \end{pmatrix},&\\
    H=\frac{1}{\sqrt{2}}\begin{pmatrix}
           1 & 1\\ 
           1 & -1
         \end{pmatrix}\quad\textrm{and}\quad
    S=\begin{pmatrix}
           1 & 0\\ 
           0 & i
         \end{pmatrix}.&
\end{split}
    \label{eq:gates}
\end{equation}

\noindent The effect of these gates can be seen by applying them to an arbitrary qubit state as given in Eq.~\ref{eq:psi}. For example,

\begin{equation}
\begin{split}
    Z\ket{\psi}=\begin{pmatrix}
       1 & 0\\ 
       0 & -1
     \end{pmatrix}\cdot \begin{pmatrix}
       \cos\left(\frac{\theta}{2}\right)\\ 
       e^{i \phi}\sin\left(\frac{\theta}{2}\right)
     \end{pmatrix} = \begin{pmatrix}
       \cos\left(\frac{\theta}{2}\right)\\ 
       -e^{i \phi}\sin\left(\frac{\theta}{2}\right)
     \end{pmatrix}\\
     = \cos\left(\frac{\theta}{2}\right) \ket{0} - e^{i \phi}\sin\left(\frac{\theta}{2}\right) \ket{1}.
\end{split}
\end{equation}

\noindent This gate, the $Z$ gate, applies what is known as phase-flip on the qubit, rotating its state $180\deg$ around the $z$-axis of the Bloch sphere. Equivalently the $X$ gate applies what is known as bit-flip on the qubit, rotating its state $180\deg$ around the $x$-axis of the Bloch sphere. These are quantum equivalents of classical logic operations. The $H$ or Hadamard gate is uniquely quantum gate. If applied to the $\ket{0}$ state the following happens

\begin{equation}
    H\ket{0} =\frac{1}{\sqrt{2}}\begin{pmatrix}
           1 & 1\\ 
           1 & -1
         \end{pmatrix}\cdot \begin{pmatrix}
           1 \\ 
           0
         \end{pmatrix}=\begin{pmatrix}
           \frac{1}{\sqrt{2}} \\ 
           \frac{1}{\sqrt{2}}
         \end{pmatrix}=\frac{1}{\sqrt{2}}\ket{0} + \frac{1}{\sqrt{2}}\ket{1}=\ket{+}.
\end{equation}

\noindent This gate has taken the classical state $\ket{0}$ on the north-pole of the Bloch sphere and rotated it down to the equator of the Bloch sphere, leaving the qubit in a superposition state referred to as the $\ket{+}$ state. 

This example is of some simple and useful single qubit gates, however there is no limit on the type of rotation on the Bloch sphere a single qubit gate can do. Any single qubit gate can be applied to take you from any one point on the Bloch sphere to the other. The only limit on what a good gate can do is keep qubit normalized. A gate $U$ satisfies this condition if it is a \textit{unitary}, i.e.

\begin{equation}
    U\cdot U^{\dagger} = \mathbb{I}
    \label{eq:unitary_condition}
\end{equation}

\noindent where $\mathbb{I}$ is the identity matrix, a diagonal matrix where all the elements are 1, and $^\dagger$ denotes the \textit{hermitian conjugate} of the gate. 

The true power of quantum computers, however, lies in the strange consequences of gates that act on two or more qubits. When describing a system with two qubits $\ket{\Psi}$ a vector of length $2^{n=2}=4$ is constructed from the \textit{tensor product} (denoted by $\otimes$) of the two single qubit vectors $\ket{\psi_1}$ and $\ket{\psi_2}$

\begin{equation}
    \ket{\Psi}=\ket{\psi_1}\otimes \ket{\psi_2}=\begin{pmatrix}
           \alpha \\ 
           \beta
         \end{pmatrix}\otimes\begin{pmatrix}
           \gamma \\ 
           \delta
         \end{pmatrix} = \begin{pmatrix}
           \alpha \cdot \gamma \\ 
           \alpha \cdot \delta \\
           \beta \cdot \gamma \\ 
           \beta \cdot \delta \\
         \end{pmatrix}.
    \label{eq:tensor}
\end{equation}

\noindent Therefore a gate that acts on both qubits at the same time, a two-qubit gate, is given by a $4\times 4$ matrix. A typical example of two-qubit gate is the controlled-not or $CNOT$ gate 

\begin{equation}
    CNOT = \begin{pmatrix}
           1 & 0 & 0 & 0\\ 
           0 & 1 & 0 & 0\\
           0 & 0 & 0 & 1\\
           0 & 0 & 1 & 0\\
         \end{pmatrix}.
\end{equation}

\noindent Note that all two-qubit gates including the $CNOT$ gate obey the same unitary condition as described in~\ref{eq:unitary_condition}. On classical inputs, the $CNOT$ gate does a bit-flip operation on the second qubit depending on the state of the first. However, when acting on qubits in superposition states, interesting things can happen. For example, if a $CNOT$ gate is applied to a two-qubit system with the first qubit in the $\ket{+}$ superposition state and the second in the $\ket{0}$ state written as

\begin{equation}
    \ket{+}\otimes\ket{0}=\frac{1}{\sqrt{2}}\begin{pmatrix}
           1 \\ 
           1
         \end{pmatrix}\otimes\begin{pmatrix}
           1 \\ 
           0
         \end{pmatrix} = \frac{1}{\sqrt{2}}\begin{pmatrix}
           1 \\ 
           0 \\
           1 \\
           0
         \end{pmatrix}
\end{equation}

\noindent the following happens

\begin{equation}
\begin{split}
    CNOT \ket{+}\otimes\ket{0} = \begin{pmatrix}
           1 & 0 & 0 & 0\\ 
           0 & 1 & 0 & 0\\
           0 & 0 & 0 & 1\\
           0 & 0 & 1 & 0\\
         \end{pmatrix} \cdot  \frac{1}{\sqrt{2}}\begin{pmatrix}
           1 \\ 
           0 \\
           1 \\
           0
         \end{pmatrix} \\
         = \frac{1}{\sqrt{2}}\begin{pmatrix}
           1 \\ 
           0 \\
           0 \\
           1
         \end{pmatrix} = \ket{\Phi_+}.    
\end{split}
\end{equation}

\noindent The reason why the state $\ket{\Phi_+}$ is interesting is that mathematically it can no longer be written as the tensor product of two single qubit states as in Eq.~\ref{eq:tensor}. This means that two qubits are now intrinsically linked or entangled. 

In the density matrix notation, to calculate the effect of a gate the matrix must be pre-multiplied by the gate must as well as post-multiplied by its Hermitian conjugate. This becomes evident, when describing the density matrix as the inner product of a column (ket) and a row (bra) vector, i.e.

\begin{equation}
\tilde{\rho} = U\cdot\rho\cdot U^\dagger = U\cdot\ket{\psi}\bra{\psi}\cdot U^\dagger.
\label{eq:density_mat_mult}
\end{equation}

\subsubsection{Unity Implementation}
In the Unity environment, five distinct single quantum gates are implemented to perform operations on the qubits. Since each single-qubit gate is a $2\times2$ matrix, to apply, for example, the $X$ gate to the second qubit of a 2 qubit scene, the following must be applied to the total $2^2\times2^2$ density matrix 

\begin{equation}
U =  \mathbb{I} \otimes X
\end{equation}

\noindent The single-qubit gates may then be applied to the total density matrix as shown in~\ref{eq:density_mat_mult}. Each gate is represented as a matrix made up of complex 32-type values and is implemented as described in function \ref{eq:gates}.





\subsection{Wavefunction}

The quantum wavefunction provides the mathematical description of a quantum state, encapsulating its evolution and probabilistic nature. The wavefunction carries information including probability densities, energy eigenstates, and superposition~\cite{klein_probabilistic_2020}.
\subsubsection{Unity Implementation} 
\begin{figure}[t]
    \centering
    \includegraphics[width=0.8\linewidth]{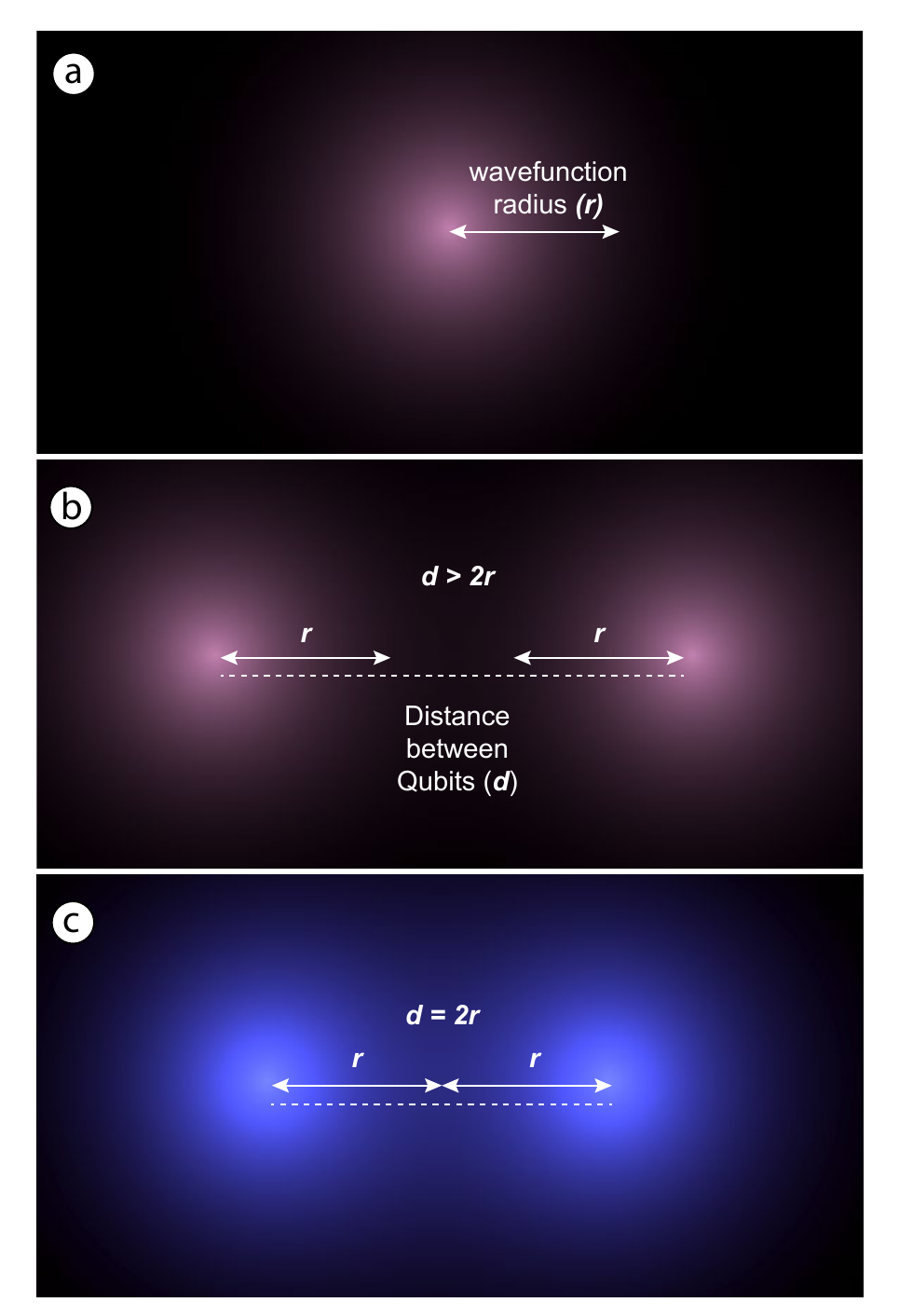}
    \caption{(a) Single qubit's wavefunction. The wavefunction is visible on the top and bottom of each qubit (b) Two qubits' wavefunction that does not overlap (c) Overlapping two qubits' wavefunctions: this condition becomes the initial condition of the entanglement and the wavefunction's color changes accordingly for the visual indication}
    \label{fig:wave}
\end{figure}

The 2D ground state of the wavefunction with a radius ($r$) for each qubit is computed using a shader programming in Unity. The code encodes the complex wavefunction as real and imaginary parts and evolves them over time using a discrete approximation of the Schrödinger equation in a fragment shader. In the XR scene, the wavefunction floating on the bottom and the top of each qubit becomes an indicator of the initial condition of the entanglement. Figure~\ref{fig:wave} shows the wavefunction in different scenarios. For instance, when the qubits are not entangled, the wavefunctions stay with a pink color indicating the unentangled state (figure~\ref{fig:wave}-b). At the moment the qubits' wavefunctions overlap (figure~\ref{fig:wave}-c), the qubits become entangled and the wavefunction turns to a purple indicating the entanglement is established.

%

\subsection{Entanglement}

Without entanglement quantum computers are just probabilistic versions of classical computers. With entanglement, quantum computers can potentially perform operations and calculations to solve problems that cannot be solved by classical computers in useful time. However, not all entangled states are equally entangled. The degree to which two or more qubits are entangled can be quantified by a measure known as the \textit{entanglement entropy}. Such measures of entanglement are derived from the density matrix of two or more qubits. Consider the arbitrary two-qubit state $\ket{\Psi}$

\begin{equation}
    \ket{\Psi} = \begin{pmatrix}
           \alpha \\ 
           \beta \\
           \gamma \\
           \delta
         \end{pmatrix}
    \label{eq:2_qubit_vector}
\end{equation}

\noindent which can be used to build a $4\times4$ density matrix

\begin{equation}
\begin{split}
    \rho=\ket{\Psi}\bra{\Psi} = \begin{pmatrix}
           \alpha \\ 
           \beta \\
           \gamma \\
           \delta
         \end{pmatrix}\cdot(\alpha^*,\beta^*,\gamma^*,\delta^*)\\
         =\begin{pmatrix}
           |\alpha|^2 & \alpha\cdot\beta^* & \alpha\cdot\gamma^* & \alpha\cdot\delta^*\\ 
           \beta\cdot\alpha^* & |\beta|^2 & \beta\cdot\gamma^* & \beta\cdot\delta^*\\
           \gamma\cdot\alpha^* & \gamma\cdot\beta^* & |\gamma|^2 & \gamma\cdot\delta^*\\
           \delta\cdot\alpha^* & \delta\cdot\beta^* & \delta\cdot\gamma^* & |\delta|^2\\
         \end{pmatrix}
\end{split}
\end{equation}

\noindent The reason why density matrix notation is useful is that even if the state $\ket{\Psi}$ is entangled and therefore cannot be written as the tensor product of two single qubit states $\ket{\psi_1}\otimes\ket{\psi_2}$, the density matrix of the entangled state $\rho$ can still be separated into the tensor product of two single qubit density matrices 

\begin{equation}
    \rho = \rho_1 \otimes \rho_2.
\end{equation}
\noindent It is from these separated entangled density matrices the second R\'enyi entropy form of the entanglement entropy can be calculated as
\begin{equation}
    S_2(\rho_1) = -\ln\left(\Tr[\rho_1^2] \right)
\end{equation}

\noindent The value of $S_2(\rho_1)$ is a useful metric of how entangled the qubit $\rho_1$ is to other qubits in the system that can vary from $0$ for a disentangled system

\begin{equation}
    S_2(\ket{0}\bra{0})=-\ln\left(\Tr\left[\begin{pmatrix}
           1 & 0\\ 
           0 & 0
         \end{pmatrix}^2\right] \right)= -\ln\left(1\right)=0
    \label{eq:Min_Entanglement}
\end{equation}

\noindent to a maximally entangled system 

\begin{equation}
    S_2(\rho_{\text{max}})=-\ln\left(\Tr\left[\begin{pmatrix}
           \frac{1}{2} & 0\\ 
           0 & \frac{1}{2}
         \end{pmatrix}^2\right] \right)= \ln\left(2\right).
    \label{eq:Max_Entanglement}
\end{equation}

\noindent These examples are the two extremes of the how entangled a qubit can be, the degree of entanglement is a sliding scale however, and so a qubit can have any entanglement entropy between these two values.

\begin{equation}
    \rho_{123} = \rho_1 \otimes \rho_2 \otimes \rho_3=\rho_{12} \otimes \rho_3=\rho_{1} \otimes \rho_{23}=\rho_{13} \otimes \rho_2
\end{equation} 

The more entangled a qubit is, the less information is held locally within the qubit, and the more said information becomes \textit{delocalized}. This can be seen when attempting to place the state of an entangled qubit on the Bloch sphere. From~\ref{eq:Cartesian_Density_Matrix}, it is evident that the radius of the Bloch sphere describing a state $\rho$ is
\begin{equation}
    r=\sqrt{u^2+v^2+w^2}=\sqrt{(\rho_{00}-\rho_{11})^2+4\rho_{01}\rho_{10}}.
\end{equation}

\noindent This radius ranges from $r=1$ for when the qubit is completely disentangled form all other qubits, such as when $\rho=\ket{0}\bra{0}$ like in~\ref{eq:Min_Entanglement}, to $r=0$ for when the qubit is maximally entangled, such as $\rho=\rho_{\text{max}}$ like in~\ref{eq:Max_Entanglement}. Therefore, a maximally entangled single-qubit state cannot be drawn on a Bloch sphere as its information has become completely delocalized.

\subsubsection{Exchange Interaction}

There are many different physical processes by which two or more qubits may become entangled. Here we will focus on emulating a naturally occurring entanglement process known as exchange between two spins. In semiconductor based quantum computers, qubits are given by an electrons the spin degree of freedom, a natural symmetry of a single electron in a small magnetic field, that can be described as either spin-up $\ket{\uparrow}=\ket{0}$ or spin-down $\ket{\downarrow}=\ket{1}$. As two spins are controllably brought within close proximity to each other, they begin to exchange their information or spin-state as a function of time and the strength of their exchange interaction (how close they are to each other). By controlling how much time the spins are allowed under exchange relative to how strongly they are interacting the two spin qubits may be controllably entangled with one another. As such, to understand how the quantum system behaves as a function of time, we must consider the \textit{time dependent Schr\"odinger equation}.

In quantum mechanics, the a quantum system is described by a Hamiltonian $H$, which in the proceeding discussion can be given as a matrix or \textit{operator} such that

\begin{equation}
    H\ket{\psi_i} = E_i \ket{\psi_i}
\end{equation}

\noindent where $\ket{\psi_i}$ is a wavefunction of a natural quantised state of the system described by $H$ known as an \textit{eigenstate} and $E_i$ is the energy of the eigenstate $\ket{\psi_i}$. The time dependent Schr\"odinger equation has the form

\begin{equation}
    H(t)\ket{\psi(t)} = i \hbar \frac{\partial}{\partial t} \ket{\psi(t)}
\end{equation}

\noindent where $\ket{\psi(t)}$ is the time dependent state of the quantum system and $\hbar$ is \textit{Planck's constant}, a fundamental universal constant that sets the energy scales of quantum systems. However, for our simulation purposes, $\hbar$ is unnecessary, and so we will assume $\hbar=1$, which is common practice in quantum mechanics and is referred to as swapping to \textit{natural units}. If the Hamiltonian of the system is time independent, i.e. $H(t)=H$ then the following solution to the time dependent Schr\"odinger equation exists

\begin{equation}
    \ket{\psi(t)} = U(t) \ket{\psi(t=0)} = e^{-i t H} \ket{\psi(t=0)}
    \label{eq:TDSE_Sol}
\end{equation}

\noindent where $U(t)$ is the operator that imparts the time evolution of the system and $\ket{\psi(t=0)}$ is the initial state, or the state at time $t=0$ of the system. This is what we will assume to describe the exchange interaction of two qubits in our simulation.

To derive the effect of exchange interaction of qubits, the following Hamiltonian known as the Heisenberg model can be used

\begin{equation}
    H_{\text{spin chain}} = \sum_{i=1}^{N-1} \frac{J_i}{4} \bm{\sigma}_i\cdot \bm{\sigma}_{i+1}.
\end{equation}

\noindent Specifically, this Hamiltonian describes a 1D chain of $N$ spins, where $J_i$ is the exchange interaction strength, given as an energy, between spins labeled $i$ and $i+1$. Finally, $\bm{\sigma}$ is the vector of Pauli matrices for each spin defined as

\begin{equation}
\begin{split}
        \bm{\sigma}&=\{\sigma_x,\sigma_y,\sigma_z\}\\ 
        \sigma_x = \begin{pmatrix}
           0 & 1\\ 
           1 & 0
         \end{pmatrix} \quad \sigma_y &= \begin{pmatrix}
           0 & -i\\ 
           i & 0
         \end{pmatrix} \quad \sigma_z = \begin{pmatrix}
           1 & 0\\ 
           0 & -1
         \end{pmatrix}.
\end{split}
\end{equation}

\noindent In the simplest case of two interacting spins, the Heisenberg model can be written as

\begin{equation}
    H_{\text{2 spins}} = \frac{J}{2} \begin{pmatrix}
           \frac{1}{2} & 0 & 0 & 0\\ 
           0 & -\frac{1}{2} & 1 & 0\\ 
           0 & 1 & -\frac{1}{2} & 0\\ 
           0 & 0 & 0 & \frac{1}{2}
         \end{pmatrix},
\end{equation}

\noindent which can be simplified as

\begin{equation}
    \tilde{H}_{\text{2 spins}} = H_{\text{2 spins}} - \frac{J}{4} \mathbb{I}_4   = \frac{J}{2} \begin{pmatrix}
           0 & 0 & 0 & 0\\ 
           0 & -1 & 1 & 0\\ 
           0 & 1 & -1 & 0\\ 
           0 & 0 & 0 & 0
         \end{pmatrix}.
\end{equation}

\noindent Such a transformation, adding or subtracting a Hamiltonian by some chosen constant multiped by the identity matrix $\mathbb{I}_4$, is commonplace, as it does not change the physics of the Hamiltonian, and serves only to add or subtract an immeasurable \textit{global phase} to the system. With this simplified form of the Hamiltonian of two interacting spins, the time-dependent form of the effect of the exchange interaction derived from Eq.~\ref{eq:TDSE_Sol} is given as

\begin{equation}
    U_{\text{2 spins}} (t) = e^{-i t \tilde{H}_{\text{2 spins}}} =\begin{pmatrix}
           1 & 0 & 0 & 0\\ 
           0 & e^{\frac{i J t}{2}} \cos \frac{J t}{2} & -i e^{\frac{i J t}{2}} \sin \frac{J t}{2} & 0\\ 
           0 & i e^{\frac{i J t}{2}} \sin \frac{J t}{2} & e^{\frac{i J t}{2}} \cos \frac{J t}{2} & 0\\ 
           0 & 0 & 0 & 1
         \end{pmatrix}.
    \label{eq:Ent_gate}
\end{equation}

\noindent $U_{\text{2 spins}} (t)$ will serve as a time-dependent entangling gate in our XR quantum computer.

Since Eq.~\ref{eq:Ent_gate} describes the exchange interaction of two spin qubits, there are a couple of interesting things to note about the gate. Firstly, if two spins are aligned, i.e. $\ket{\uparrow\uparrow}=\ket{00}$ or $\ket{\downarrow\downarrow}=\ket{11}$, exchanging their information does nothing, and the following happens when the gate is applied

\begin{equation}
    U_{\text{2 spins}} (t) \begin{pmatrix}
           \alpha \\ 
           0 \\
           0 \\
           \delta
         \end{pmatrix}=\begin{pmatrix}
           \alpha \\ 
           0 \\
           0 \\
           \delta
         \end{pmatrix}
\end{equation}

\noindent However, when the gate is applied to spins that are anti-aligned, i.e. $\ket{\uparrow\downarrow}=\ket{01}$ or $\ket{\downarrow\uparrow}=\ket{10}$, the following happens

\begin{equation}
    U_{\text{2 spins}} (t) \begin{pmatrix}
           0 \\ 
           \beta \\
           \gamma \\
           0
         \end{pmatrix}=\begin{pmatrix}
           0 \\ 
           \beta \cos \frac{J t}{2} -i \gamma \sin \frac{J t}{2} \\
           \gamma \cos \frac{J t}{2} +i \beta \sin \frac{J t}{2} \\
           0
         \end{pmatrix}
\end{equation}

\noindent which as a function of time, oscillator between a non-entangled state and a maximally entangled state. The frequency of these oscillations is given by the parameter $J$. In a spin-qubit system, this parameter is dependent on how much overlap there is between the wavefunctions of the two interacting electrons. As such, to emulate this in our system, the parameter $J(\Delta_r)$ is a function of the distance $\Delta_r$ between the qubits as they are moved around the environment by the user(s). Specifically the following fit is used

\begin{equation}
    J(\Delta_r)=\frac{J_{max}}{2}(1+\tanh \frac{\Delta_{\text{cutoff}}}{2}-\Delta_r)
\end{equation}

\noindent where $J_{max}$ is the maximum value of J and $\Delta_{\text{cutoff}}$ is the cutoff distance at which there is no allowed interaction between the qubits. Such a function allows for a smooth, natural feeling transition from no interaction between qubits, to a maximum strong qubit-qubit interaction.

Lastly, the entangling gate is applied to the density matrix of the system similarly to single qubit gates, i.e.

\begin{equation}
    \rho (t) = U_{\text{2 spins}} (t) \rho_0 U^\dagger_{\text{2 spins}} (t).
\end{equation}

\subsubsection{Applying the Heisenberg Model in Unity}
The implementation begins with the construction of a two-spin Hamiltonian, parameterized by the interaction strength $J$. Using the matrix exponential \( U(t) = e^{-iHt} \), the unitary evolution operator is derived, enabling the simulation of quantum state transitions over time (function \ref{eq:Ent_gate}). The matrix exponential is approximated using a truncated series expansion for computational efficiency.

To simulate the dynamic interactions, qubits in a spatial environment are associated with Unity game objects. Their pairwise distances are calculated, and interactions are determined based on a threshold distance. The coupling constant $J$ is scaled proportionally to the proximity of qubits, reflecting the physical principle that interaction strength decreases with distance. The function is described as follow: 

\begin{equation}
J(\Delta_r) = \frac{J_{\max}}{2}
\biggl(
  1 + \tanh\!\frac{\Theta_d}{2} - \Delta_r
\biggr),
\end{equation}
\noindent{where $\Theta_d$ is the threshold distance and \(0 \le \Delta_r \le \Theta_d\), $J_{\max}$ = 1. The threshold distance $\Theta_d$ may vary depending on the user's environment. In this case, we set $\Theta_d$ = 5f as a sample value. The time evolution of the quantum system is applied to the density matrix using the unitary operator, ensuring the quantum state evolves correctly while preserving entanglement.}

\begin{figure} 
    \centering
    \includegraphics[width=1\linewidth]{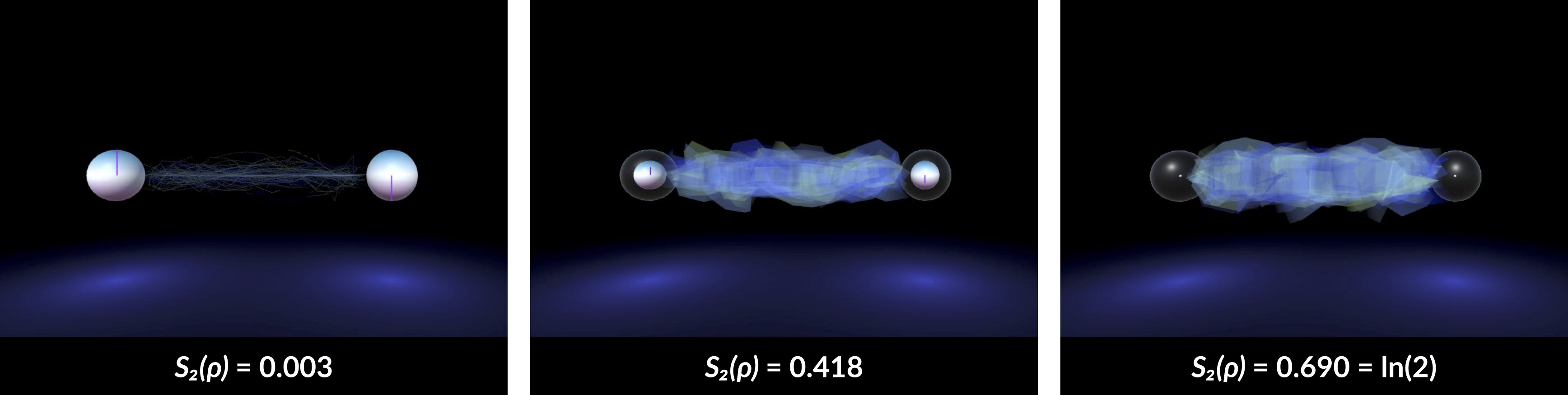}
    \caption{Examples of entanglement visualization varying by entanglement entropy $S_2(\rho)$. The higher $S_2(\rho)$ creates a more chaotic and vivid entanglement pattern, while decreases the Bloch sphere radius-emphasizing the delocalization of quantum information.}
    \label{fig:entanglement}
\end{figure}

To visualize the entanglement between two qubits in the proposed XR experience, an arc of oscillating colorful lines are drawn between the two qubits. The entanglement entropies of the system are used to parameterize the intensity of the colors and oscillations, to communicate the different degrees of entanglement between different pairs of qubits to the user. Specifically, the following pairwise entanglement entropy parameter is used:

\begin{equation}
    \tilde{S}_{ij}=S_2(\rho_i)+S_2(\rho_j)-S_2(\rho_{ij}).
\end{equation}

\noindent Here the difference between sum of the entanglement entropies experienced by two qubits $i$ and $j$ in isolation and the entanglement entropy of both qubits relative to all other qubits being simulation $S_2(\rho_{ij})$ gives an accurate representation of the distribution of entanglement. Examples of entanglement visualization varying by entanglement entropy are shown in figure \ref{fig:entanglement}.

\subsection{Measurements}
A vital ingredient for a quantum computer is measurement. When the state of a qubit is measured, the information held in the qubit collapses from whatever quantum state it is in, to one of the outcomes of the chosen measurement probabilistically. There are many different types of measurements considered in quantum information science, but here focus will be made on the most common: the single qubit $z$ measurement. This measurement takes a single qubit and collapses it to either the $\ket{0}$ or $\ket{1}$ state, forcibly setting the qubit into one of these two states, with probability given by the initial state of the qubit. Measuring all the qubits involved in a quantum computing task this way is typically how the outcome of the task is determined, and can be thought of as translating the quantum information to classical information. A single qubit $z$ measurement can be written as follows

\begin{equation}
    \Pi_z (s) = \frac{\mathbb{I}_2 + (-1)^s \sigma_z}{2}
\end{equation}

\noindent where $\Pi_z (s)$ is known as a projector along the $z$-axis states of the qubits Bloch sphere and $s=\{0,1\}$ is the outcome of the measurement. Explicitly, the two possible measurement projectors can be written as 

\begin{equation}
    \Pi_z (0) = \ket{0}\bra{0}=\begin{pmatrix}
        1 & 0 \\
        0 & 0
    \end{pmatrix}
    \label{eq:Pi_0}
\end{equation}

\noindent for when $s=0$ and the qubit is collapses to the $\ket{0}$ state and

\begin{equation}
    \Pi_z (1) = \ket{1}\bra{1}=\begin{pmatrix}
        0 & 0 \\
        0 & 1
    \end{pmatrix}
    \label{eq:Pi_1}
\end{equation}

\noindent for when $s=1$ and the qubit is collapses to the $\ket{1}$ state. It is important to remember that quantum measurements are probabilistic, meaning, if the state $\ket{\psi}=\{\alpha,\beta\}=\{\cos \theta/2, e^{i \phi}\sin \theta/2\}$ then the probability of measuring the state as $\ket{0}$ depends on how much that state's superposition is in the $\ket{0}$ state, which is given by the magnitude of the vector component $|\alpha|^2=\cos^2 \theta/2$. Equally, the probability of measuring a single qubit as $\ket{1}$ is $|\beta|^2=\sin^2 \theta/2$. Note that, as per the normalization condition, $|\alpha|^2+|\beta|^2=1$ meaning that when a qubit is measured in this way, the outcome must always be ether $\ket{0}$ or $\ket{1}$. In the density matrix formalism where $\rho=\ket{\psi}\bra{\psi}$, these probabilities are calculated as follows

\begin{equation}
    P_s = \Tr\left(\Pi_z (s)\cdot\rho\right)
\end{equation}

\noindent were if $s=0$

\begin{equation}
    P_0 = \Tr\left(\Pi_z (0)\cdot\rho\right)=\bra{0}\rho\ket{0}
\end{equation}

\noindent and if $s=1$

\begin{equation}
    P_1 = \Tr\left(\Pi_z (1)\cdot\rho\right)=\bra{1}\rho\ket{1}.
\end{equation}

Qubit measurements are a vital part of the quantum information toolbox, and when appropriately combined with single and entangling gates, can serve not only as a way of reading out the states of a quantum computer, but also as an additional method of controlling or manipulating the information in important applications of quantum computers such as measurement-based quantum computing, quantum teleportation and quantum error correction.

\subsubsection{Unity Implementation} 
Users can initiate qubit measurement by holding a qubit and pressing X/A button on the Meta Quest controllers. The partial trace for the corresponding qubit is used to determine the probabilistic measurement outcome. The partial trace function isolates this single qubit information from the simulation of every other qubit in the scene. From the density matrix describing a single qubit, as shown in~\ref{eq:Single_Density_Matrix}, the value $\rho_{00}$ is equivalent to the probability of the state collapsing to $\ket{0}$ when measured, whilst the value $\rho_{11}$ is equivalent to the probability of the state collapsing to $\ket{1}$. A pseudo-random real number between $0$ and $1$ is then generated and used to select what state the qubit will be collapsed or reset to, as weighted by the measurement probabilities. Once determined, the full density matrix is multiplied by a matrix $\Pi_z(0)/\sqrt{\rho_{00}}$ as shown in~\ref{eq:Pi_0} or $\Pi_z(1)/\sqrt{\rho_{11}}$ as shown in~\ref{eq:Pi_1} depending on the state collapsed to, similarly to how the single qubit gates are implemented. These additional factors $1/\sqrt{\rho_{00}}$ and $1/\sqrt{\rho_{11}}$ ensure that the total density matrix simulated in the scene remains normalized.

\section{System Validation}
To ensure the functional correctness and interaction fidelity of QIXR, we conducted a comprehensive technical validation process. The evaluation focused on the accuracy of qubit and Bloch sphere behavior, quantum gate transformations, entanglement representations, measurement operations, and the system’s real-time performance within an immersive virtual environment. The test methodology was informed by established approaches to XR system evaluation~\cite{billinghurst_survey_2015}~\cite{bowman_3d_2011}. The video demonstration is available at this URL \href{https://www.myunginlee.com/qixr}{https://www.myunginlee.com/qixr}

\subsection{Technical Validation}
Ten structured test cases were developed and executed to validate the system’s core functionalities. Table \ref{tab:validation-tests} summarizes a subset of representative test cases. 

\begin{table*}[ht]
\centering
\begin{tabularx}{\linewidth}{@{} X X X X @{}}
\toprule
\textbf{Objective} & \textbf{Input/Action} & \textbf{Expected Output} & \textbf{Validation Method} \\
\midrule
Validate Bloch sphere & Initiate qubits in $\ket{0}$ and $\ket{1}$ state & Bloch vector aligns with +Z ($\ket{0}$) and –Z  ($\ket{1}$) & Visual observation \\
Validate Hadamard gate transformation & Apply H gate to qubit in $\ket{0}$ state & Bloch vector rotates to +X axis (superposition) & Visual observation + internal state log \\
Detect entanglement  & Move qubits within threshold proximity & Entanglement arc appears based on entropy; Bloch radius oscillates; audio/color feedback & Visual/Audio cues + internal log \\ 
Verify measurement & Press X/A button while holding entangled qubit & Bloch vector aligns to Z-axis; entanglement breaks & Bloch sphere + state vector log \\
Test performance under load & Simulate rapid gate use and entanglement & Maintain frame rate $\geq$ 80 FPS & OVR Metric Tools + in-headset observation \\
\bottomrule
\end{tabularx}
\caption{Representative system validation test cases.}
\label{tab:validation-tests}
\end{table*}


\subsubsection{Qubit, Bloch sphere, and quantum gate transformation} 

We validated the correctness of quantum gate operations—Hadamard (H), Pauli-X (X), Pauli-Z (Z), and Phase-S (S)—by applying each to qubits initialized in known basis states (e.g., $\ket{0}$ or $\bra{1}$) and comparing the results to theoretical expectations. Each qubit is visualized as a Bloch sphere, where the quantum state is encoded via its polar and azimuthal angles ($\theta$, $\phi$). Gate operations result in real-time Bloch sphere rotations, offering intuitive feedback on state changes. 

\begin{figure} 
    \centering
    \includegraphics[width=\linewidth]{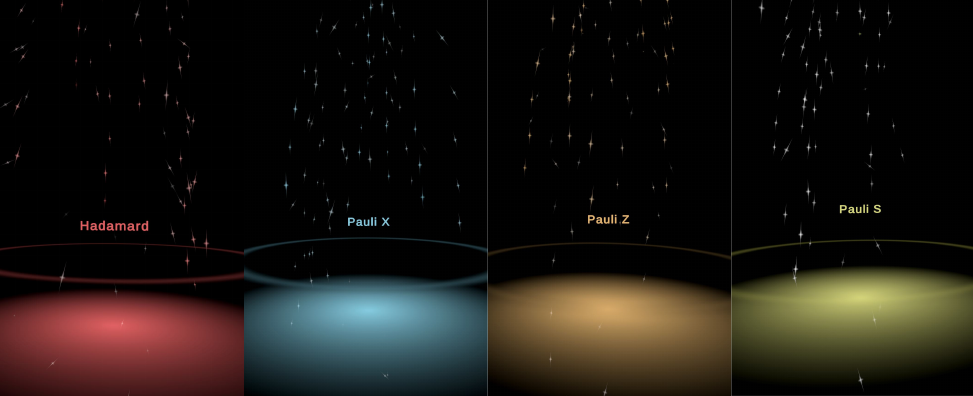}
    \caption{Implemented quantum gates in the system: Hadamard, Pauli-X, Pauli-Z, Phase-S}
    \label{fig:gates}
\end{figure}

For example, applying a Hadamard gate to a qubit in $\ket{0}$ state correctly rotated the Bloch vector from the +Z axis to the +X axis, representing a superposition state $(\ket{0} + \bra{1})/\sqrt{2}$. Similarly, applying the Pauli-X gate correctly flipped the qubit from $\ket{0}$ to $\ket{1}$, corresponding to a vector inversion along the Z axis. These visual transformations were confirmed by internal log of the qubit’s partial trace, ensuring consistency with expected unitary transformations. This combination of Bloch sphere visual feedback and backend state verification confirmed the accurate implementation of Bloch sphere and quantum gate behaviors. Quantum gates visualization can be found in figure \ref{fig:gates}.

\subsubsection{Entanglement} 

We evaluated the entanglement mechanic by bringing two qubits into close proximity and observing the trigger of visual and auditory indicators. When the qubits entered a defined spatial threshold, an animated, color-coded arc appeared between them, accompanied by a finger-snap auditory cue. While the two qubits are entangled, qubits' color change was observed, and the entanglement interaction generates string-like auditory cues with pitch variations corresponding to the qubits’ spin angular values.

\begin{figure}[b]
    \centering
    \includegraphics[width=\linewidth]{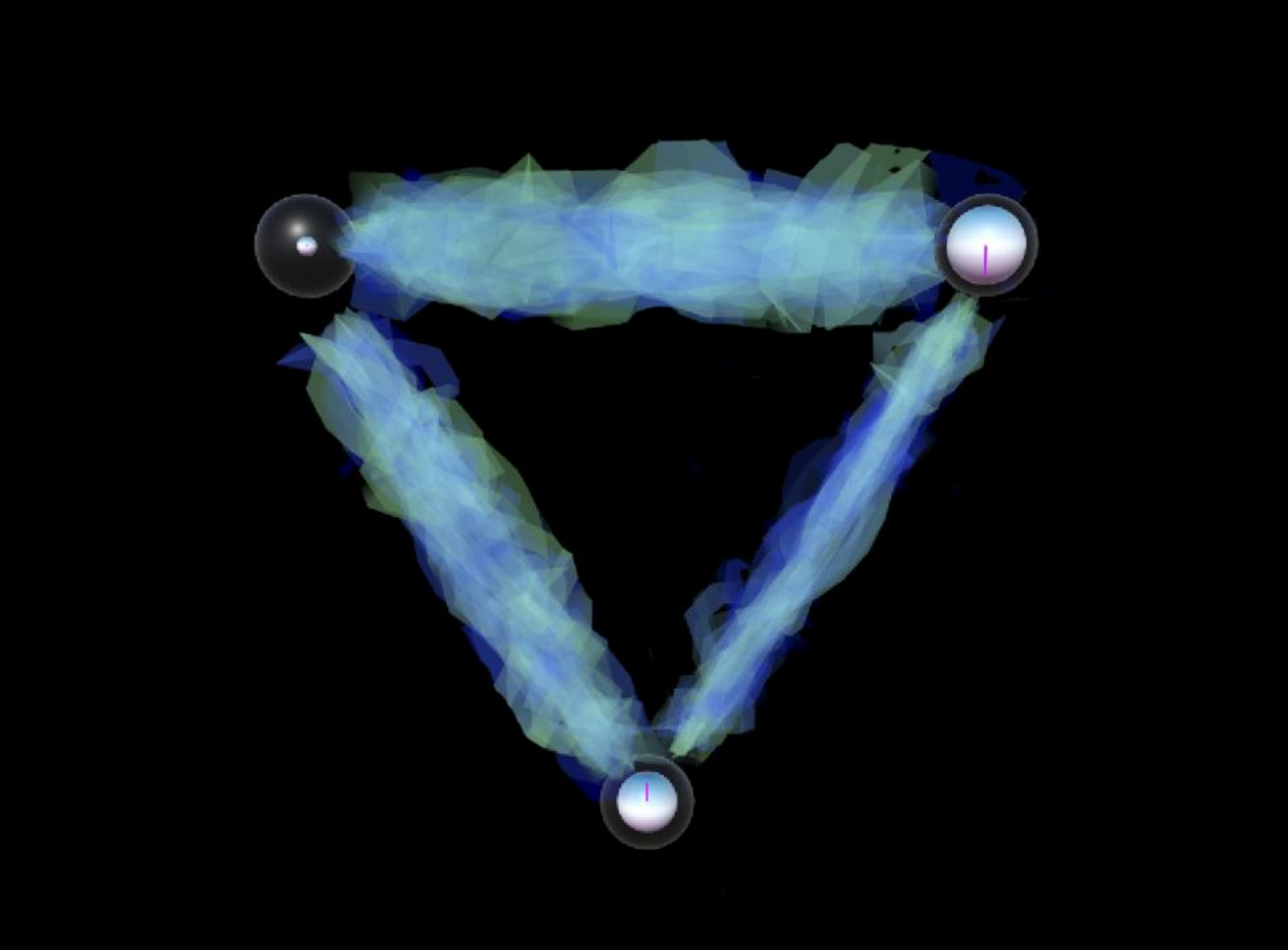}
    \caption{An exemplary entanglement of three qubits}
    \label{fig:3-qubit}
\end{figure}

The entanglement is further emphasized through oscillating wave patterns within the arc, with their amplitude and frequency modulating based on the real-time entanglement entropy calculation. The entanglement strength is visually represented using color intensity and dynamic distortions in the arc, with highly entangled qubits produce a more vibrant, fluctuating connection, while weakly entangled pairs exhibit faint and subtle linking effects. As entanglement increases, each qubit begins to shrink in visual prominence, while the connecting arc becomes more apparent--emphasizing the delocalization of quantum information. The qubit shrinking effect is quantitatively governed by the Bloch radius parameter, which dynamically adjusts the Bloch sphere’s size based on the degree of entanglement. While the current version supports two qubits scenarios, we are developing a multi-user and multi-qubit version for further release. Figure \ref{fig:3-qubit} shows an example of the entanglement of three qubits. 

\begin{figure}[t]
    \centering
    \includegraphics[width=\linewidth]{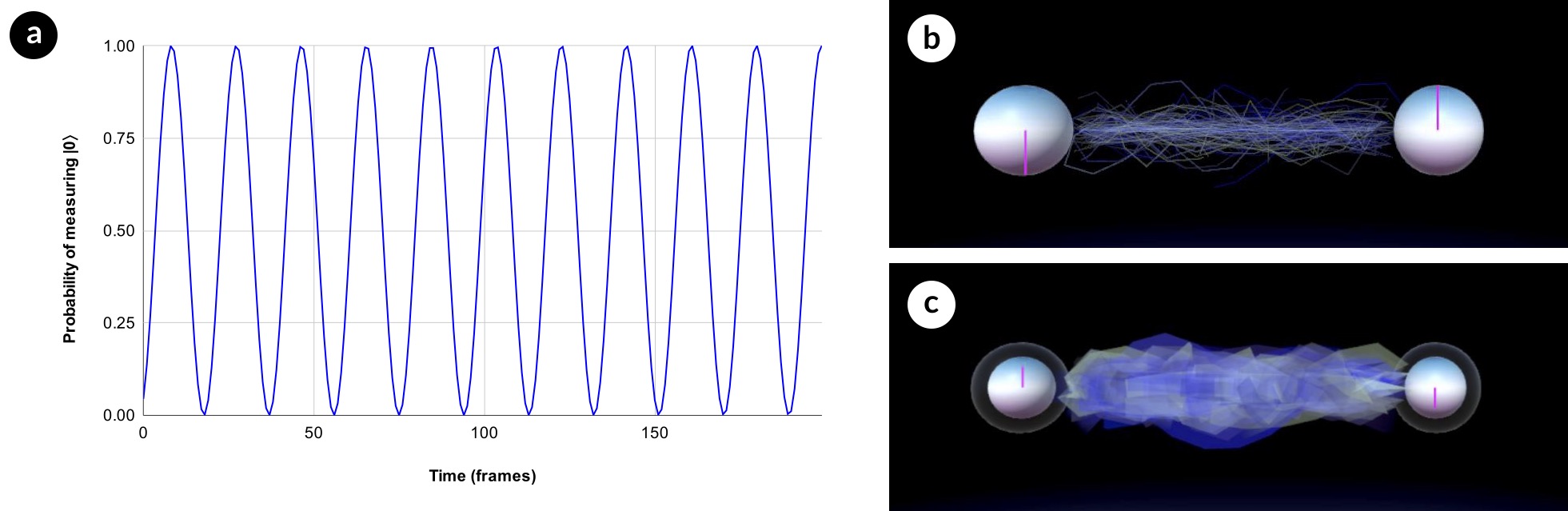}
    \caption{a) Plot of the probability of measuring the $\ket{0}$ state, as a function of time, of a qubit initialized to the $\ket{1}$ state undergoing an exchange-interaction with a qubit starting in the $\ket{0}$ state. b), c) Visualization of entanglement between two qubits initialized at $\ket{1}$ and $\ket{0}$, undergoing exchange.}
    \label{fig:osillation}
\end{figure}

Additionally, we logged the density matrix of the entangled qubit pair at each frame, along with the partial trace of each individual qubit, to compare against expected theoretical values. The partial trace results were plot the evolution of each qubit’s reduced state over time. These graphs exhibit the expected oscillatory behavior, confirming the accuracy of the entanglement dynamics. An example is shown in figure \ref{fig:osillation}, where qubit initialized in the $\ket{1}$ state exhibits oscillations between 1 and 0 in its reduced density matrix, consistent with theoretical predictions, when interacting with a qubit initialized to the $\ket{0}$ state.

\subsubsection{Measurement}
The system supports both instantaneous and time-frozen quantum measurement modes. In the standard mode, users may measure a qubit’s state by pressing the X or A button on the controller. If the qubit is entangled, this action collapses its state and breaks the entanglement. This is reflected in the realignment of its Bloch vector, the disappearance of the connecting arc, and the halting of the Bloch sphere’s oscillatory motion. In time-freeze mode, users may hold the button to pause the simulation temporally, enabling inspection of the quantum system without immediately collapsing the entangled state. Both modes were validated through internal state logging and observational testing.

\subsubsection{Real-time Performance}
To evaluate usability and system responsiveness, we tested all core interaction pathways, including qubit grabbing, manipulation, gate placement, and entanglement triggering. All interactions were performed in a natural spatial range and were free of latency, jitter, or frame loss. Testing under peak interaction conditions (simultaneous entanglement rendering, gate transformations, and measurement operations) showed consistent frame rates above 80 FPS on the Meta Quest 3 headset. This confirms the system’s capacity to maintain real-time responsiveness and comfort within the virtual stimulation.

\subsection{Expert Interview}
We conducted structured expert interviews with five participants (4 males, 1 female), all of whom hold advanced degrees in quantum physics and have experience researching quantum science. Participants reported having little to no prior VR experience. The evaluation combined a live in-person system demonstration with Meta Quest 3 with a post-experience survey. Our demonstration took place in a quiet university building and lasted around 15 minutes.

All five experts completed the survey. Responses indicated a high level of agreement across all items, particularly in areas related to the accuracy of Bloch sphere, quantum gates and entanglement implementations. The experts also provided informal verbal feedback following the survey. Several highlighted the use of extended reality as a novel and engaging approach to visualizing quantum phenomena. Suggestions for future improvements included incorporating time-dependent gates and modeling decoherence to enhance the system's functional depth. The detailed quantitative and qualitative analysis of this survey will be presented in our future work.

\section{Discussion}
One of the most notable benefits of the proposed system is \textbf{the ability to represent complex physical phenomena audiovisually and interactively}. Traditional physics education often relies on abstract mathematical representations and two-dimensional illustrations, which can be challenging for learners to fully grasp. In contrast, designated multimodal experiences enhance learning by engaging multiple sensory channels, which fosters deeper cognitive processing, better memory encoding, and more robust conceptual understanding~\cite{sigrist_augmented_2013}. The QIXR's artistically inspired and scientifically accurate visualization renders the core components in the scene, and these components are fully interactable using a gesture-based XR controller. The proposed sonification enables invisible or temporal quantities to be perceived efficiently. Audio can efficiently deliver temporal and spectral information without physical disturbance~\cite{riecke_self-motion_2012}. Sonification of the components such as the exchange interaction based on the interaction strength presents a coherent multimodal interpretation of the quantum phenomenon. At the same time, spatial audio delivers dynamic spin information from multiple qubits and entanglement scenarios~\cite{hwang_effect-_2018}. Such multisensory experience has potential to allow users to engage with quantum mechanics in a way that mirrors real-world experiences. 

Furthermore, such experiential learning can \textbf{enhance retention and motivation.} The users can manipulate variables, observe real-time changes, and receive immediate feedback. This hands-on approach fosters curiosity and deeper cognitive engagement compared to passive learning methods~\cite{andersen_fostering_2023}. Gamification elements in XR simulations, such as interactive challenges and scenario-based learning, further increase motivation and knowledge retention~\cite{lampropoulos_virtual_2024}.


\section{Conclusion}
While quantum mechanics is often perceived as an abstract and unintuitive field due to its counterintuitive principles, QIXR leverages XR to bridge this gap by providing an interactive and immersive environment where users can directly engage with quantum phenomena. Through our XR-based system, participants manipulate qubits, explore superposition and entanglement, and gain an embodied understanding of quantum mechanics.

Our approach demonstrates that XR can serve as a powerful tool for visualizing and intuitively understanding complex quantum principles. By allowing users to experience quantum interactions through direct manipulation and spatial awareness, we provide a novel method for teaching quantum computing concepts beyond traditional mathematical formalisms.

Looking ahead, we aim to expand the system’s capabilities by improving the precision of quantum state representations, incorporating more complex quantum operations, and enhancing multi-user collaboration. By refining the interaction model and integrating additional educational elements, we envision QIXR as a scalable and engaging platform for both education and research in quantum computing.

\section{Acknowledgments}
This work is supported by the University of Maryland's Immersive Media Design program and Quantum ArtsAMP grant.


\bibliographystyle{abbrv-doi-hyperref}

\bibliography{01_references-myungin}

\end{document}